\documentclass[preprint,aps,nofootinbib,preprintnumbers,amsmath,amssymb,11pt]{revtex4}

\usepackage{amsmath}
\usepackage{amssymb}
\usepackage{setspace}

\usepackage{amsfonts}
\usepackage{graphicx}
\usepackage{dcolumn}
\usepackage{bm}
\usepackage{natbib}
\newcommand{\half}{\frac{1}{2}}
\newcommand{\beq}{\begin{equation}}
\newcommand{\eq}{\end{equation}}
\newcommand{\bea}{\begin{eqnarray}}
\newcommand{\ea}{\end{eqnarray}}
\newcommand{\p}{\partial}
\newcommand{\nn}{\nonumber}
\begin{document}
\preprint{MIFPA-10-17}
\preprint{DAMTP-2010-43}

\title{\LARGE  Correlation Functions and Hidden Conformal Symmetry of Kerr Black Holes\\}

\author{\vspace*{1 cm}\large
Melanie Becker$^\dagger$, Sera Cremonini$^{\dagger,\S}$, Waldemar Schulgin$^\dagger$}
\email{mbecker,sera,schulgin@physics.tamu.edu}
\affiliation{\vspace*{0.5 cm}$^\dagger$ George and Cynthia Mitchell Institute for Fundamental Physics and Astronomy,
Texas A\&M University, College Station, TX 77843--4242, USA\\}
\affiliation{$^\S$ Centre for Theoretical Cosmology, DAMTP, CMS,\\
University of Cambridge, Wilberforce Road, Cambridge, CB3 0WA, UK}

\begin{abstract}
\vspace*{1 cm}
Extremal scalar three-point correlators in the near-NHEK geometry of Kerr black holes have recently been shown to agree
with the result expected from a holographically dual non-chiral two-dimensional conformal field theory.
In this paper we extend this calculation to extremal three-point functions of scalars in a general Kerr black hole
which need not obey the extremality condition $M=\sqrt{J}$.
It was recently argued that for low frequency scalars in the Kerr geometry there is a dual conformal field theory description which
determines the interactions in this regime. Our results support this conjecture.
Furthermore, we formulate a recipe for calculating finite-temperature retarded three-point correlation functions which is 
applicable to a large class of (even non-extremal) correlators, and discuss the vanishing of the extremal couplings.
\end{abstract}

\vspace*{3 cm}
\maketitle

\newpage
\tableofcontents

\def\thesection{\arabic{section}}
\def\thesubsection{\arabic{section}.\arabic{subsection}}
\numberwithin{equation}{section}


\section{Introduction}

In the original formulation of the Kerr/CFT correspondence \cite{Guica:2008mu}, quantum gravity in the NHEK 
(near-horizon extreme Kerr) geometry was conjectured to have a dual description in terms of a \emph{chiral}, 
left-moving two-dimensional CFT with central charge $c_L=12J$, where $J$ denotes the black hole angular momentum.
This conjecture was motivated by the fact that the asymptotic symmetry group of the NHEK geometry was shown to be
one copy of the conformal group.
Support for the proposal emerged from the agreement between the macroscopic Bekenstein-Hawking entropy and the
microscopic entropy of the dual CFT, whose computation relied crucially on knowledge of the central charge $c_L=12J$
of the Virasoro algebra \cite{Guica:2008mu}.
The presence of only a chiral half of the CFT can be explained by the fact that at extremality the black hole horizon
rotates at the speed of light. Since both edges of the forward light-cone coincide as the horizon is approached,
all physical excitations are forced to rotate chirally with the black hole.

A generalization of the original Kerr/CFT conjecture to the near-NHEK geometry (near-extremal, near-horizon Kerr)
emerged shortly thereafter \cite{Bredberg:2009pv}.
The near-extremal case allows for some energy above extremality -- the forward light cones no longer coincide, and 
right-moving excitations are now possible, in addition to left-movers.
The theory has been conjectured \cite{Bredberg:2009pv} to be dual to a 2D \emph{non-chiral} CFT with $c_L=c_R=12J$.
Even though finding appropriate boundary conditions allowing for both left-movers and right-movers has proven elusive,
several studies offer support for the conjecture \cite{Bredberg:2009pv,Matsuo:2009sj,Castro:2009jf}.
In addition to the matching of microscopic and macroscopic entropies, it is also supported
by the agreement between finite-temperature two-point correlation functions -- on the CFT side -- and black hole scattering
amplitudes -- on the gravity side -- for frequencies close to the superradiant bound 
(see \cite{Bredberg:2009pv,Cvetic:2009jn,Chen:2010ni}).

At the level of correlation functions, a more systematic check was provided only recently by the calculation of
extremal three-point functions of scalars in the near-NHEK geometry \cite{Becker:2010jj}, arising from 
interaction terms of the form $\sim  \lambda \, \Phi_{h_1} \Phi_{h_2} \Phi_{h_3}$.
On the CFT side, these are finite-temperature correlators of the form 
$\langle {\cal O}_{h_1} {\cal O}_{h_2} {\cal O}_{h_3} \rangle$.
Such correlators are ``extremal'' when the conformal weights $h_i$
of the operators ${\cal O}_{h_i}$ dual to the scalar fields obey $h_3=h_1+h_2$.
Such a restriction dramatically simplifies the form of the CFT three-point function, reducing it to a product of two two-point functions,
while still providing a strong test for the conjecture.

In an interesting new development \cite{Castro:2010fd}, the wave-equation for a \emph{low-frequency} scalar 
in the background of a general Kerr black hole has been shown to exhibit an $SL(2,R)_L \times SL(2,R)_R$ symmetry.
In addition to the requirement of low frequencies, \emph{i.e.} $\omega M \ll 1$,
the conformal symmetry arises from the near-region of Kerr, specified by $r \ll 1/\omega$.
Since these conditions don't place any restrictions on the temperature (for very small frequencies, $r$ becomes arbitrarily large),
the analysis of \cite{Castro:2010fd} provides additional evidence
for the  validity of the Kerr/CFT conjecture for general values of mass and angular momentum.
In particular, it justifies using the Cardy formula to compute the microscopic entropy, since it allows one to take 
the temperatures to be large compared to the central charges.
Scattering computations of low-frequency scalars in the near-region of Kerr
were again shown to reproduce two-point correlation functions on the CFT side
\cite{Castro:2010fd, Chen:2010xu}, in close analogy with the near-NHEK scattering calculations.
Finally, we emphasize that thus far the $SL(2,R)_L \times SL(2,R)_R$ is only understood as a symmetry of the equation of motion,
and \emph{not} an isometry of the metric -- hence the name ``hidden'' conformal symmetry.
While this conformal symmetry acts locally on the space of solutions, it is obstructed globally
by periodic identification of the azimuthal angle $\phi$.
For recent extensions of \cite{Castro:2010fd} see 
\cite{Rasmussen:2010xd,Chen:2010zw,Chen:2010xu,Wang:2010qv,Chen:2010as,Krishnan:2010pv}.

In this note we would like to extend our previous near-NHEK computation of 
scalar three-point correlation functions to the low-frequency case analyzed in \cite{Castro:2010fd}, 
in the background of a general Kerr black hole.
Although we restrict our attention to the particular case of extremal correlators, our computation can also 
be applied to non-extremal conformal weights, as we explain in Section \ref{Recipe}. 
The calculation proceeds along the lines of \cite{Becker:2010jj}.
On the gravity side, we will see that the dominant contribution to the three-point function
comes from a term which contains a divergent factor $\propto \frac{1}{h_3-h_1-h_2}$. 
In particular, the gravity result reproduces the expected CFT correlators, 
\beq
\lambda \, \int \Phi_{h_1} \Phi_{h_2} \Phi_{h_1+h_2} 
\sim \lambda \; \frac{1}{h_3-h_1-h_2} \; \langle {\cal O}_{h_1} {\cal O}_{h_2} {\cal O}_{h_1+h_2} \rangle \, ,
\eq 
and tells us that the coupling $\lambda$ of the cubic interaction should vanish for extremal correlators, 
\beq
\lambda_{\;\text{extremal}} \propto h_3-h_1-h_2 \, ,
\eq
in direct analogy with standard AdS/CFT studies of extremal correlators \cite{Lee:1998bxa,D'Hoker:1999ea,Skenderis:2006uy}.
The vanishing of the extremal coupling is expected -- and dictated -- from the structure of conformal anomalies 
(see e.g. \cite{Skenderis:2006uy,Petkou:1999fv})
in theories that admit a Coulomb branch, as we explain in Section \ref{Vanishing}.
This check provides one more piece of evidence for the existence of a dual, non-chiral CFT description
of general Kerr black holes.
We emphasize that our three-point function calculation can be directly generalized to a variety of backgrounds, 
as we explain in Section \ref{Recipe}, and is therefore in some sense ``universal.''
Because of its relative simplicity, in Section \ref{Recipe} we will outline its main features, so that it can 
be easily adopted in related contexts.

This paper is organized as follows. In Section \ref{HCS} we review some highlights of the hidden conformal symmetry found in 
\cite{Castro:2010fd}.
Section \ref{Calc} is devoted to the calculation of the extremal three-point correlation function, for low-frequency scalars 
in a general four-dimensional Kerr black hole.
In Section \ref{Vanishing} we explain why the coupling of extremal correlators is expected to vanish, and its relation to 
the conformal anomaly.
Section \ref{Recipe} offers a summary of the main ingredients of our three-point function calculation (touching 
on \cite{Becker:2010jj} as well as this note), including comments on its applicability to non-extremal correlators.
We conclude with a discussion of our results, open problems and work in progress.

\section{Hidden Conformal Symmetry}
\label{HCS}

It has been recently
observed in \cite{Castro:2010fd}
that \emph{at low frequencies} the wave equation for a scalar field incident on a Kerr black hole exhibits
two-dimensional conformal symmetry, for \emph{generic} non-extreme values of the mass, $M \neq \sqrt{J}$.
In particular, \cite{Castro:2010fd} has been able to recast the scalar wave equation (for sufficiently low frequencies,
$\omega M \ll 1$ and in the region where $r\ll \frac{1}{\omega}$) in terms of the generators of an $SL(2,R)_L \times SL(2,R)_R$
symmetry. This symmetry, however, is not an isometry of the geometry (unlike the NHEK case, where there was an 
$SL(2,R)\times U(1)$ isometry), but just a statement about what certain scalar modes see while scattering off the Kerr black hole.
Thus, this is an instance where the solution space has the requisite conformal symmetry, but the space
on which the field propagates does not. Let's outline how this works, following the discussion of \cite{Castro:2010fd}.

Consider a massless scalar field incident on a Kerr black hole,
\begin{equation}
\Phi(t,r,\theta,\phi) = e^{-i\omega t+i m\phi} \phi (r,\theta).
\end{equation}
It is well-known that the wave equation $\Box \Phi =0$ in the Kerr background separates.
Thus, for a field of the form
\begin{equation}
\Phi(t,r,\theta,\phi) = e^{-i\omega t+i m\phi} R(r) S(\theta),
\end{equation}
the wave equation for the \emph{radial} wavefunction is given by
\begin{equation}
\label{FullRadialEqn}
\Biggl[ \partial_r \Delta \partial_r + \frac{(2Mr_+\omega-am)^2}{(r-r_+)(r_+-r_-)}-\frac{(2Mr_-\omega-am)^2}{(r-r_-)(r_+-r_-)}
+\Bigl(r^2+2M(r+2M)\Bigr)\omega^2 \Biggr]R(r) =K_l R(r) \, ,
\end{equation}
with
\begin{equation}
\Delta = r^2+a^2-2Mr\, , \quad a=\frac{J}{M} \, , \quad r_\pm = M \pm \sqrt{M^2-a^2} \, .
\end{equation}
The radial equation can be greatly simplified by dropping the last term,
which is ${\cal O}(\omega^2)$.
More precisely, in the limit where the wavelength of the scalar excitation is much larger than the curvature radius,
\begin{equation}
\omega  \ll \frac{1}{M} \, ,
\end{equation}
and in the ``near region'' $r\ll 1/\omega$, the angular wave equation reduces to the standard laplacian on a two-sphere $S^2$,
the
separation constant becomes $K_l =l(l+1)$ and the radial equation (\ref{FullRadialEqn}) simplifies to
\begin{equation}
\label{nearWaveEqn}
\Bigl[ \p_r \Delta \p_r + \frac{(2Mr_+\omega-am)^2}{(r-r_+)(r_+-r_-)}-\frac{(2Mr_-\omega-am)^2}{(r-r_-)(r_+-r_-)}
 \Bigr]R(r) = l (l+1) R(r) \, .
\end{equation}
Note that the near region $r\ll 1/\omega$ is \emph{not} the near-horizon region of the Kerr black hole.
In fact, when $\omega$ is small enough, $r$ can become arbitrarily large, in contrast to the NHEK geometry.
The crucial observation of \cite{Castro:2010fd} is that the near-region radial wave equation (\ref{nearWaveEqn})
can be recast in terms of appropriately identified $SL(2,R)_L \times SL(2,R)_R$ generators:
\begin{equation}
\label{casimir}
H^2 \Phi = \bar{H}^2 \Phi = l(l+1)\Phi \, ,
\end{equation}
where $H^2 = \half(H_1 H_{-1} + H_{-1}H_1)-H_0^2$, and an analogous expression for $\bar{H}^2$, denote the
$SL(2,R)$ Casimirs. We refer the reader to \cite{Castro:2010fd} for the explicit form of the generators.

The solution to the radial wave equation for low frequency scalars $\omega M \ll 1$ and incoming boundary conditions 
at the horizon has been known for a long time\footnote{For the computation of the absorption cross-section see. 
e.g. \cite{Maldacena:1997ih,Cvetic:1997xv}.}.
In the notation of \cite{Castro:2010fd} it takes the form
\beq
\label{radialwvfn}
R(r)=\Bigl(\frac{r-r_+}{r-r_-}\Bigr)^{\frac{-2 i M r_+(\omega-m\Omega)}{r_+-r_-}} (r-r_-)^{-l-1}
\; F\Bigl(\alpha,\beta;\gamma;\frac{r-r_+}{r-r_-}\Bigr) \, ,
\eq
with
\beq
\label{parameters}
\alpha \equiv 1+l- \frac{4iM}{r_+-r_-}(M\omega-r_+m\Omega) \; , \quad
\beta \equiv 1+l-2iM\omega \; , \quad
\gamma \equiv 1-\frac{4iMr_+}{r_+-r_-}(\omega-m\Omega)\, .
\eq
In the asymptotic regime $r\gg M$ of the near region $r\ll\frac{1}{\omega}$ it reduces to the simple form
\begin{equation}
\label{eqA}
R(r)\sim Ar^l+B r^{-l-1}.
\end{equation}
The coefficients are given by
\begin{equation}
\label{Acoeff}
A=\frac{\Gamma\left(1-i\frac{4Mr_+}{r_+-r_-}(\omega-m\Omega)\right)\Gamma(1+2l)}{\Gamma(1+l-2iM\omega)
\Gamma\left(1+l-\frac{i 4 M^2}{r_+-r^-}\omega+\frac{i4Mr+\Omega}{r_+-r_-}m\right)}\, ,
\end{equation}
where the horizon angular velocity is $\Omega=\frac{a}{r_+^2}\sim\frac{a}{2Mr_+}$,  and
\begin{equation}
\label{Bcoeff}
B=\frac{\Gamma(-2l-1)\Gamma\left(1-i\frac{4Mr_+}{r_+-r_-}(\omega-m\Omega)\right)}{\Gamma(-l-2iM\omega)
\Gamma\left(-l-\frac{i 4 M^2}{r_+-r_-}\omega+\frac{i4Mr+\Omega}{r_+-r_-}m\right)} \, .
\end{equation}
As we will see in the next section, this asymptotic expansion will play a key role in the computation of
extremal three-point functions.

\section{Extremal Three-Point Correlators}
\label{Calc}

Extremal three-point functions of scalars in the near-NHEK geometry were shown to agree with the corresponding finite
temperature conformal field theory correlators in \cite{Becker:2010jj}.
Here we would like to perform an analogous computation of extremal three-point correlators, but in the general Kerr black hole geometry, under the assumption that the scalar fields have low frequency, $\omega \ll 1/M$.
The calculation can be done in a manner analogous to \cite{Becker:2010jj}, so we shall be brief.
For a more detailed treatment of the calculation we refer the reader to \cite{Becker:2010jj}.

On the conformal field theory side, we are interested in computing finite temperature three-point correlators of the form
$\langle {\cal O}_{h_1} {\cal O}_{h_2} {\cal O}_{h_3} \rangle$, where the $h_i$
denote the conformal dimensions of the operators. Since the conjectured dual CFT is non-chiral, we will be dealing with two sectors -- right (left) movers at temperature $T_R$ ($T_L$).
On the gravity side, we will be looking at three-point functions of scalar fields $\Phi_h$
dual to the operators we are interested in.
As shown in \cite{Becker:2010jj}, the finite temperature CFT three-point correlator\footnote{As we will explain 
in Section \ref{Recipe}, this is a \emph{retarded} correlator.}
can be obtained, on the gravity side, by the bulk integral over three bulk-to-boundary propagators,
\begin{equation}\label{integral}
\langle {\cal O}_{h_1}(\vec x_1) {\cal O}_{h_2}(\vec x _2) {\cal O}_{h_3}(\vec x_3) \rangle
\sim \int_{r_+}^{r_{\rm {eff}}}
d \vec x \, dr \, \sqrt{-g} K_{h_1}(r,\vec x; \vec x_1) K_{h_2}(r,\vec x;  \vec x_2) K_{h_3}(r,\vec x;  \vec x_3) \, .
\end{equation}
The radial integral runs from the black hole horizon $r_+$ to the location $r_{\rm {eff}}$ of the ``effective boundary''
where the dual CFT lives. For the case of \cite{Castro:2010fd}, the boundary is specified by $M \ll r \ll  \frac{1}{\omega}$.
Note that by making $\omega$ arbitrarily small, one can push the boundary to infinity.
We will discuss the prescription (\ref{integral}) -- and its applicability to the finite temperature case -- in the next section.
For now we just mention that the bulk-to-boundary propagator is constructed so that it obeys incoming wave
boundary conditions at the black hole horizon.

Computing the three-point function using the exact solution to the wave equation, i.e. the full hypergeometric
function $F(\alpha,\beta;\gamma;\frac{r-r_-}{r-r_+})$ is challenging, and moreover is not needed for extremal correlators.
In fact, the wavefunction can be approximated in various regions, and can then be
used in a straightforward way for the computation of n-point correlation functions.
In \cite{Becker:2010jj} we approximated the wavefunction near the horizon, near the boundary, and in the intermediate 
region, by making appropriate expansions in $\tau_H \sim r_+ - r_-$.
After computing the resulting three-point integral we took the conformal weights to be extremal, 
$h_3=h_1+h_2$, which simplified the result significantly.  
However, \cite{Becker:2010jj} does contain an (approximated) expression for the three-point function which is valid for 
generic, non-extremal conformal weights, under the assumption that $\tau_H$ is small.

Here we would like to adopt an alternative derivation of the three-point correlator, which is more concise than that 
given in \cite{Becker:2010jj}, and is particularly appropriate to the special case of extremal correlators.
We should note, however, that this derivation fails for non-extremal correlators -- for generic conformal weights 
the reader should follow the procedure of \cite{Becker:2010jj}, which we spell out in the next section.

We start by expanding the radial wavefunction (\ref{radialwvfn}) (which coincides with the momentum-space propagator \cite{Becker:2010jj}) 
around the boundary,
which is specified by $r\gg M$ as long as the condition $r\ll\frac{1}{\omega}$ is satisfied.
By taking $r\gg r_+-r_-=2\sqrt{M^2-a^2} \sim 2M$, the hypergeometric function can be approximated by the infinite series
\begin{eqnarray}
\lim_{r \gg M} \;
 {_2F_1}\left(\alpha,\beta;\gamma;\frac{r-r_-}{r-r_+}\right)&=&\left(\frac{r}{r_--r_+}\right)^{\alpha+\beta-\gamma}\left[\frac{
 \Gamma(\alpha+\beta-\gamma)\Gamma(\gamma)}{\Gamma(\alpha)\Gamma(\beta)}+{\cal{O}}\left(\frac{r_+-r_-}{r}\right)\right]\nonumber\\
&&+\left[\frac{\Gamma(\gamma)\Gamma(\gamma-\alpha-\beta)}{\Gamma(\gamma-\alpha)\Gamma(\gamma-\beta)} +{\cal{O}}
\left(\frac{r_+-r_-}{r}\right)\right] \, ,
\end{eqnarray}
allowing one to rewrite the wavefunction as
\begin{eqnarray}
\label{wavef}
 R(r)&=&  A\, r^l \, (r_--r_+)^{-2l-1}  + {\cal O}\Big((r_--r_+)^{-2l}r^{l-1}\Big)\nonumber\\
 &+&  B\,r^{-1-l}+{\cal O}\Big((r_--r_+)r^{-2-l}\Big) \, .
\end{eqnarray}
The coefficients $A$, $B$ are precisely the ones given in (\ref{Acoeff}) and (\ref{Bcoeff}).
At any value of $r$ along  the integration region from the outer horizon to the effective boundary the series is convergent.
Note that we are \emph{not} truncating the series in (\ref{wavef}), although we are using a suggestive notation for the 
subleading terms ${\cal {O}}(\ldots)$, for reasons which will become clear.

Typically if one is interested only in the boundary behavior of the wavefunction -- as in two-point correlation 
function calculations -- the expansion (\ref{wavef}) is truncated,
and one keeps only the leading normalizable and non-normalizable modes, as in (\ref{eqA}).
However, we are interested in using (\ref{wavef}) to compute the three-point function integral, 
which requires integrating over the entire bulk, and we should therefore
keep as many terms in the expansion as possible.
Eventually, when the radial coordinate becomes close to the horizon, $r\sim r_+$,
this approximation to the wavefunction will break down\footnote{Note that in the wavefunction expansion we implicitly
assumed that $\frac{r-r_+}{r_--r_+}\approx \frac{r}{r_--r_+}$.}, and we will be forced to approximate $R(r)$ in a different manner.
It is straightforward to approximate the full solution near the horizon $r_+$, as well as when $r\sim r_+ - r_-$.
For explicit details of the approximations in the near-NHEK region of Kerr we refer the reader to \cite{Becker:2010jj}.
It turns out that for the types of correlators we are interested in, the contribution to the bulk integral comes
\emph{from the boundary} and will contain a divergent term, which dominates all remaining (finite) terms, 
giving an exact expression for the three-point function integral. 
Thus, because of this divergence we don't need the (finite) terms arising from the regions $r \gtrsim r_+$.

The solution of the radial wave equation (\ref{wavef}) matches the \emph{momentum-space} 
bulk-to-boundary propagator \cite{Becker:2010jj}.
It is more convenient to perform our computation in momentum space. 
Using the relation between the momentum and coordinate space propagator
\begin{equation}
\label{bulktoboundary}
K(r,t^\prime,\phi^\prime;t, \phi) = \int dm \int d \omega \; \tilde K(r,m,\omega) \; e^{-im(\phi-\phi^\prime)}
e^{i \omega (t-t^\prime)}\, ,
\end{equation}
and the expression for the three-point correlator in coordinate space
\beq
\langle O(t_1,\phi_1)\, O(t_2,\phi_2)\, O(t_3,\phi_3)\rangle
\sim \int d\phi^\prime  dt^\prime  dr \, \sqrt{-g} K_1(r,t^\prime,\phi^\prime;t_1,\phi_1)
K_2(r,t^\prime,\phi^\prime;t_2,\phi_2) K_3(r,t^\prime,\phi^\prime;t_3,\phi_3)\, ,
\eq
it is a simple task to a obtain a corresponding expression for the correlator in momentum space:
\begin{eqnarray}\label{intmomentum}
V_3^{m.s.} &=& \langle O(m_1,\omega_1)\, O(m_2,\omega_2)\, O(m_3,\omega_3)\rangle \nn \\
&=& \delta(m_1 + m_2 + m_3) \, \delta(\omega_1+ \omega_2 + \omega_3)
\int_{r_+}^{r_{\rm{eff}}} dr \, \sqrt{-g} \;\tilde K_1 \tilde K_2 \tilde K_3 \, .
\end{eqnarray}
Taking into account the normalization appropriately \cite{Becker:2010jj} and using the relation 
between the angular momentum and the conformal weight, which can be read off from (\ref{casimir}) 
\beq
h=l \, ,
\eq
we find the final expression for the momentum-space bulk-to-boundary propagator:
\begin{eqnarray}\label{prop}
\tilde K(r,m,\omega) &=& r^{h} + {\cal O}\Big((r_--r_+)r^{h-1}\Big)\nonumber\\
 &+& \frac{{ B}}{{ A}}\,r^{-h-1}\,(r_--r_+)^{2h+1}+{\cal O}\Big((r_--r_+)^{2h+2}r^{-2-h}\Big) \, .
\end{eqnarray}
Inserting (\ref{prop}) into the three-point integral (\ref{intmomentum}) and performing the integration term by term will give us
one divergent term
and infinitely many finite terms (part of which can be identified as contact terms).
For the full details of the calculation we again refer the reader to \cite{Becker:2010jj}.
Here we note that the terms one gets are of the form:
\begin{eqnarray}
V_3^{m.s.} &=&\int_{r_+}^{r_{\rm {eff}}} \, \sqrt{-g} \, \tilde{K}_{h_1}\tilde{K}_{h_2}\tilde{K}_{h_3}=
\Biggl[{\rm{ contact\ terms}}+ \frac{B_1B_2B_3}{A_1A_2A_3}\frac{r^{-h_1-h_2-h_3+1}}{1-h_1-h_2-h_3} (r_--r_+)^{2h_1+2h_2+2h_3+3}\nonumber\\
&&+  \frac{B_1 B_2}{A_1 A_2} \; \frac{r^{h_3-h_1-h_2}}{h_3-h_1-h_2}\ (r_--r_+)^{2h_1+2h_2+2} \;
+ \frac{B_2 B_3}{A_2 A_3} \;  \frac{r^{h_1-h_2-h_3}}{h_1-h_2-h_3} \ (r_--r_+)^{2h_2+2h_3+2} \;\nn \\
&& +\frac{B_1 B_3}{A_1 A_3}  \; \frac{r^{h_2-h_1-h_3}}{h_2-h_1-h_3}\; (r_--r_+)^{2h_1+2h_3+2}+\ldots \Biggr]_{r_+}^{r_{\rm {eff}}} \, .
\end{eqnarray}
We should mention that the metric determinant in the final bulk integral is $\sqrt{-g} \sim r$.
Also, since we are working in momentum space the contact terms do not have the standard form $\delta(\vec x_i-\vec x_j)$.
We recognize them as coming from terms that only depend on a single ratio $B/A$, or from terms with no dependence on $A,B$.\footnote{For example, since the $B_1/A_1$ term will only contain dependence on $(m_1,\omega_1)$, its Fourier transform over
$(m_2,m_3,\omega_2,\omega_3)$ is trivial, and gives $\sim \delta(t_2-t_3)\delta(\phi_2-\phi_3)$.}

In the extremal limit $h_3=h_1+h_2$ the final result for the tree-point function is
\begin{equation}\label{res}
V_3^{ms}\sim (T_LT_R)^{2h_1+2h_2+2}\left(\frac{1}{h_3-h_2-h_1}\frac{{ B}_1 { B}_2}{{A}_1 {A}_2}+
{\rm {finite}}\right)\delta(m_1 + m_2 + m_3) \, \delta(\omega_1+ \omega_2 + \omega_3) \, ,
\end{equation}
where we used the relation (assuming $r_{+}\gg r_+-r_-$)
\begin{equation}
 T_LT_R=\frac{1}{(4\pi)^2}\frac{(r_++r_-)(r_+-r_-)}{r_-r_+}\approx
 \frac{1}{8\pi^2}\frac{r_+-r_-}{r_-} \, .
\end{equation}
The coefficients $A(\omega,m)$ and $B(\omega,m)$ are given by (\ref{Acoeff}) and (\ref{Bcoeff}).  
Here $A_1 \equiv A(\omega_1,m_1), A_2 \equiv A(\omega_2,m_2)$ and similarly for the $B$'s.
As in \cite{Becker:2010jj} this correlator was computed by analytic continuation from the non-extremal case and it clearly 
diverges in the extremal limit $h_3=h_1+h_2$. 
As shown in \cite{Becker:2010jj}, the three-point correlator (\ref{res}) exactly agrees with the result expected for an
extremal correlator in a finite temperature non-chiral CFT:
\bea
 \langle {\cal O}(\omega_{L_1},\omega_{R_1})  {\cal O}(\omega_{L_2},\omega_{R_2})
 {\cal O}(\omega_{L_3},\omega_{R_3}) \rangle  &\sim& \delta(\omega_{L_1}+\omega_{L_2}+\omega_{L_3}) \,
 \delta(\omega_{R_1}+\omega_{R_2}+\omega_{R_3}) \nn \\
& \times&
\langle {\cal O}(\omega_{L_1},\omega_{R_1})  {\cal O}(0,0) \rangle \;
\langle {\cal O}(\omega_{L_2},\omega_{R_2})  {\cal O}(0,0)\rangle  \nn \, ,
\ea
once we recall that each two-point function is simply given by 
$G_R(\omega_L,\omega_R) = \frac{B(\omega_L,\omega_R)}{A(\omega_L,\omega_R)}$, 
and identify the left- and right-moving frequencies appropriately:
\beq
\omega_L =\frac{2M^3}{J}\, \omega \, , \quad \quad
\omega_R =\frac{2M^3}{J}\, \omega -m \, .
\eq

The above divergence is a reflection of the vanishing of the coupling of the 
cubic interaction $\lambda \Phi^3$ in the extremal case,
\begin{equation}
\lambda\propto h_3-h_2-h_1,
\end{equation}
This ensures that the full three-point function stays finite, as we already briefly discussed.
Such a behavior of the cubic coupling for extremal correlators can be understood in the context of holographic renormalization, 
as we explain in the next section.

\section{Extremal Couplings and Conformal Anomaly}
\label{Vanishing}

As we saw in Section 3, the leading contribution to the three-point correlator in the extremal case
$h_3=h_1+h_2$  is a term of the form
\beq
\lambda \; \frac{1}{h_3-h_1-h_2} \; \frac{B_1 B_2}{A_1 A_2}\, ,
\eq
which, as we recall from \cite{Becker:2010jj}, results from a log term in disguise.
The fact that the bulk integral diverges when the conformal weights are extremal indicates that the overall coupling $\lambda$
of the interaction should vanish, as was already noted in \cite{Becker:2010jj}.

The same structure -- the vanishing of extremal couplings -- was found in the original AdS/CFT three-point calculations
(see e.g. \cite{Lee:1998bxa} and \cite{D'Hoker:1999ea}).
In that context, the couplings could be obtained by direct supergravity reduction, and were explicitly shown to vanish.
However, it was later understood that -- for theories that admit a Coulomb branch -- the vanishing of the couplings of
extremal correlators is dictated by the structure of the conformal anomalies\footnote{We are very grateful to K.
Skenderis for pointing this out to us.}.
Here we would like to briefly outline why this is so, borrowing mainly from \cite{Skenderis:2006uy}.

An important feature of the gauge/gravity correspondence is the so-called UV/IR connection,
\emph{i.e.} the standard field theoretic UV divergences are related to gravitational IR divergences due to the infinite volume of AdS.
On the gravity side, such IR divergences -- being long distance effects -- clearly come from the boundary
of the spacetime, and the procedure developed to deal with them is known as holographic renormalization
\cite{de Haro:2000xn,Bianchi:2001kw}.
At the core of the holographic renormalization scheme is the addition of appropriate counterterms, constructed
to render the action and correlation functions finite.
In particular, the computation of correlation functions can be reformulated in terms of
\emph{renormalized} 1-point functions in the presence of sources.
Specifically, for a CFT operator $O(x)$ of dimension $h$, dual to a bulk scalar field
$\Phi_h(x,r)$ in $AdS_{d+1}$, one has:
\bea
\label{1ptfn}
\langle O(x) \rangle &=& \frac{1}{\sqrt{g_0(x)}} \frac{\delta S_{ren}}{\delta \phi_0(x)}
\sim \phi_{2 h-d}(x) \, , \nn \\
\langle O(x_1)\ldots O(x_n) \rangle &\sim&
\frac{\delta\phi_{2 h-d}(x_1)}{\delta \phi_0(x_2) \ldots \delta \phi_0(x_n)}|_{\phi_0=0}  \, .
\ea
Here $S_{ren}$ denotes the renormalized on-shell action, which includes all the appropriate counterterms,
while $\phi_{2 h-d}(x)$ is a term appearing in the asymptotic expansion\footnote{The asymptotic 
expansion is performed using Graham-Fefferman coordinates
$ds^2=\frac{1}{r^2}(dr^2+g_{ij}(x,r) \,dx^i dx^j)$.}
of the scalar field $\Phi$ near the boundary:
\beq
\label{asymptoticexpansion}
\Phi_h (x,r) =r^{d-h} \, \Bigl(\phi_0 +r^2 \phi_2 + \ldots \Bigr) +
r^{h} \Bigl( \phi_{2h-d} +
\log r^2 \; \psi_{2h-d} \Bigr)
+ \ldots \; .
\eq
The coefficient $\psi_{2 h-d}$ of the log term as well as the coefficients $\phi_n$ in the asymptotic expansion
with $n<2 h -d$ are uniquely determined from the scalar field equation in terms of the source $\phi_0$.
On the other hand, $\phi_{2 h-d}$ is left undetermined by the field equation, but is specified by
the vacuum expectation value $\langle O \rangle$ of the dual operator, as shown by (\ref{1ptfn}).

Moreover, the coefficient of the log term, $\psi_{2 h-d}$, is directly related to conformal
anomalies (see e.g. \cite{Petkou:1999fv} and \cite{de Haro:2000xn}).
A crucial point for understanding the vanishing of the extremal couplings is that $\psi_{2 h-d}$ vanishes when the source
is set to zero. Thus, when $\phi_0=0$ the expansion of the scalar field should not contain any logarithmic terms.
But in the studies of extremal correlators in the AdS/CFT context (see e.g. \cite{Skenderis:2006uy}),
it was shown that such couplings arise quite generically precisely from terms in the scalar field expansion which are
logarithmic.

To sketch how this works, let's consider a particular example from the Coulomb branch of ${\cal N}=4$ SYM, where chiral primaries get a VEV.
Consider the equation of motion for the scalar field $\Phi_4$ dual to a dimension 4 operator,
\beq
\label{phi4eq}
\Box \Phi_4 = \lambda \Phi_2^2 + \ldots \; .
\eq
From the form of this equation, it is clear that $\lambda$ is the coupling of an interaction term $\sim \lambda \, \Phi_4 \Phi_2 \Phi_2$.
Moreover, since this is precisely an extremal correlator ($h_1=h_2=2$, $h_3=4$ satisfy $h_3=h_1+h_2$),
$\lambda$ is an extremal coupling.
Plugging into the equation above the near-boundary expansion for the field $ \Phi_2$ (dual to a dimension 2 operator)
\beq \Phi_2 = \langle O  \rangle \; r^2 + \ldots \eq
and, neglecting terms that are irrelevant for this analysis,  one can solve (\ref{phi4eq}) and obtain:
\beq
\Phi_4 = r^4 \log r^2 \lambda \langle O \rangle^2 + \ldots \; .
\eq
Since log terms must be absent when the sources are set to zero, the extremal coupling $\lambda$ must be forced
to vanish. This argument -- which relies crucially on the fact that log terms are proportional to sources and thus
evaluate to zero on the Coulomb branch -- can be extended to all extremal couplings.

In summary, if a CFT has a Coulomb branch, then in any holographic realization the extremal couplings should vanish.
It is plausible that a similar argument can be made for the extremal correlators that were computed herein.
It would be interesting to understand this in detail, especially given that not much is known about detailed
properties of the CFT dual to Kerr.

\section{Three-Point Correlators Recipe}
\label{Recipe}

The development of a prescription for computing generic real-time, finite-temperature n-point
correlation functions in the gauge/gravity duality has proven very challenging.
A rather simple Minkowski recipe for two-point functions has been formulated (first by \cite{Son:2002sd} and
subsequently by \cite{Herzog:2002pc}), and has become widely accepted.
However, much less explored is the case of three-point (and higher n-point) functions (see \cite{Skenderis:2008dg}
for the main work, and more recently \cite{Barnes:2010jp}).
Thus, our three-point correlation function calculation (in \cite{Becker:2010jj} for a superradiant mode in
the near-NHEK geometry of Kerr, and here for the low-frequency scalars in Kerr) serves a dual purpose, providing:
\begin{itemize}
\item
a non-trivial check
of the conjectured Kerr/CFT correspondence in the presence of both right- and left-moving excitations
\item
an explicit computation of finite temperature three-point functions for the rather non-trivial background
of a rotating black hole, relying on a
relatively simple adaptation of the Euclidean AdS/CFT prescription.
\end{itemize}

Since our computation \cite{Becker:2010jj} can be applied to a much larger class of
backgrounds than just the four-dimensional Kerr black hole, it is worth emphasizing the main ingredients:
\begin{enumerate}
\item
On the gravity side of the correspondence, we adopted the Euclidean three-point function AdS/CFT prescription,
encoded schematically in Witten's diagrams.
We reproduced (in \cite{Becker:2010jj} as well as in this paper)  the finite-temperature CFT three-point correlator by computing
the bulk integral of three bulk-to-boundary propagators:
\beq
\label{bulkint}
\langle {\cal O}_{h_1}(x)  {\cal O}_{h_2}(y)  {\cal O}_{h_3}(z) \rangle \sim
\int_{r_{horizon}}^{r_{boundary}} dr dx^\prime K_{h_1}(r,x^\prime;x) K_{h_2}(r,x^\prime;y) K_{h_3}(r,x^\prime;z) \; .
\eq
The bulk-to-boundary propagator is constructed with incoming boundary conditions at the horizon \cite{Becker:2010jj}.
Thus, what we are computing is the \emph{retarded} real-time three-point function\footnote{See 
\cite{Barnes:2010jp} for another recent implementation.}, which can be obtained
by analytic continuation in frequency space from the imaginary-time, finite-temperature correlator \cite{Kobes:1990kr}.
\item
We restricted our attention to Matsubara frequencies, for which
\beq
G_R(\omega)=G_E(i\omega_E) \, ,
\eq
\emph{i.e.} the retarded Green's function and the Euclidean two-point function coincide, provided the frequencies are
analytically continued.
Working at the Matsubara frequencies ensures that the Euclidean prescription (\ref{bulkint}), upon
analytically continuing the frequencies, yields the correct retarded real-time three-point function.
\item
Computing the three-point function (\ref{bulkint}) using the exact solution to the wave equation (e.g. in our case
the full hypergeometric function $F(a,b;,c;z)$) is challenging, and moreover is not needed.
The wavefunction can be approximated (see \cite{Becker:2010jj} for a detailed analysis) by appropriate expansions near the black hole horizon (Region I),
near the boundary (Region III), and in the intermediate region (Region II).
The bulk integral can then be estimated by patching together the contributions from the
various regions. This procedure works for non-extremal correlators as well as extremal.
The typical scalar field expansion near the boundary (focusing on the radial wavefunction) is of the form
\beq
\label{bdexp}
\Phi(r) \sim A r^{-\Delta_-} + B r^{-\Delta_+} \, .
\eq
The coefficients $A, B$ play a crucial role, for example, in the determination of the retarded two-point function:
\beq
G_R \sim \frac{B}{A} \, ,
\eq
and turn out to also play a key role in the three-point function computation \cite{Becker:2010jj}.
\item
We focused on extremal correlators $\langle {\cal O}_{h_1}  {\cal O}_{h_2}  {\cal O}_{h_3}\rangle$,
for which the sum of any two of the conformal weights equals the third, e.g. $h_3=h_1+h_2$.
In the case of extremal conformal weights several simplifications arise:
\begin{itemize}
\item
On the CFT side, the three-point correlator reduces to the product of two two-point correlators. 
Working now explicitly in momentum space, 
and denoting by $(\omega_L,\omega_R)$ the left and right-moving frequencies, we have schematically: 
\bea
\label{cft3pt}
 \langle {\cal O}(\omega_{L_1},\omega_{R_1})  {\cal O}(\omega_{L_2},\omega_{R_2}) 
 {\cal O}(\omega_{L_3},\omega_{R_3}) \rangle  &\sim& \delta(\omega_{L_1}+\omega_{L_2}+\omega_{L_3}) \,
 \delta(\omega_{R_1}+\omega_{R_2}+\omega_{R_3}) \nn \\
& \times&  
\langle {\cal O}(\omega_{L_1},\omega_{R_1})  {\cal O}(0,0) \rangle \;
\langle {\cal O}(\omega_{L_2},\omega_{R_2})  {\cal O}(0,0)\rangle  \nn \, .
\ea
\item
On the gravity side, when $h_3=h_1+h_2$ the crucial contribution to the bulk integral (\ref{bulkint})  comes from the boundary region.
Thus, it results from the wavefunction approximation (\ref{bdexp}), and takes the form
\beq
\label{div}
\sim  \frac{1}{h_3-h_1-h_2} \; \frac{B(\omega_{L_1},\omega_{R_1})}{A(\omega_{L_1},\omega_{R_1})} \frac{B(\omega_{L_2},\omega_{R_2})}{A(\omega_{L_2},\omega_{R_2})} \, .
\eq
This quantity is divergent since $h_3=h_1+h_2$.
Apart from contact terms, all other contributions to the bulk integral (coming from Regions I and II)
are \emph{finite}, and can therefore be neglected compared to (\ref{div}).
\item
Using the well-known relation $G_R \sim \frac{B}{A}$ between the asymptotic expansion of the scalar field and the retarded Green's function,
we conclude that the three-point function becomes
\beq
\sim \frac{1}{h_3-h_1-h_2} \; \langle {\cal O}(\omega_{L_1},\omega_{R_1})  {\cal O}(0,0) \rangle \;
\langle {\cal O}(\omega_{L_2},\omega_{R_2})  {\cal O}(0,0)\rangle \, ,
\eq
which is precisely of the expected form (\ref{cft3pt}), apart from the overall divergent prefactor.
\item
The divergence when $h_3=h_1+h_2$ indicates that the coupling $\lambda$ of the interaction term
$\lambda \int  \Phi_{h_1}(x) \Phi_{h_2}(y) \Phi_{h_1+h_2}(z)$ for extremal correlators should contain a vanishing prefactor,
\beq
\lambda \propto (h_3-h_1-h_2) \, ,
\eq
ensuring that the entire correlator is finite.
\end{itemize}
\item
The strategy of approximating the entire bulk integral in the various regions I, II and III is valid for extremal as well as
generic conformal weights. The advantage of the extremal computation is that the result in that case is exact, since one finds a
divergent term that dominates over all other contributions. In the case of generic, non-extremal weights, on the other hand,
the three-point function result we presented in \cite{Becker:2010jj} is only an approximation of the full three-point function.
Nonetheless, it should be a good enough approximation to capture the essential features of the CFT.
\item
Our approximation scheme and three-point function results should apply quite generically to a large class of black hole backgrounds.
The main ingredients which we needed were the knowledge of the Matsubara frequencies, and a boundary expansion the wavefunction
of the form (\ref{bdexp}), which is quite generic in backgrounds that are asymptotically $AdS$, or at least contain an
$AdS$ factor somewhere, which plays a crucial role.
\end{enumerate}

\section{Conclusions}

In this paper we have shown that the prescription developed in \cite{Becker:2010jj} for calculating finite-temperature 
extremal three-point correlation functions can be applied not only to the near-NHEK geometry, 
but also to the more recent situation \cite{Castro:2010fd} describing low-frequency scalars in a general Kerr black hole background. 
Our results provide further support for the conjectured holographic description of Kerr black holes in terms of a dual 
non-chiral conformal field theory.
The crucial ingredient that allowed us to straightforwardly apply the techniques of \cite{Becker:2010jj} to the 
present situation is the presence of an asymptotic expansion for the wavefunction of the form
\beq
\label{lasteq}
R \sim A r^{-\Delta_-} + B r^{-\Delta_+} \, ,
\eq
with $A$ and $B$ denoting, respectively, the coefficients of the leading non-normalizable and normalizable modes.
For the case of extremal correlators this boundary behavior was enough to compute the three-point function.
An asymptotic expansion of the form (\ref{lasteq}) arises generically from the near-boundary behavior of scalars in the 
background of asymptotically AdS black holes, as well as the near-horizon region of rotating Kerr black holes.
In the case considered here, it is a consequence of the $SL(2,R)_L \times SL(2,R)_R$ symmetry of the wave equation 
for a low-frequency scalar in the background of a general four-dimensional Kerr black hole -- not just in its near-horizon 
region.

We would like to claim that our prescription for calculating extremal three-point functions, which we outlined in 
Section \ref{Recipe}, applies universally to all situations 
in which an asymptotic expansion of the field of the form (\ref{lasteq}) can be found, 
and will yield an expression of the form:
$$ \lambda \; \frac{1}{h_3-h_1-h_2} \; \frac{B_1 \, B_2}{A_1 \, A_2} \, .$$
In such situations agreement with a dual conformal field theory is guaranteed by our prescription.
This is somewhat analogous to the universality of the retarded Green's function expression, 
which is also determined in a simple way in terms of $B$ and $A$:
$$G_R = B/A \, .$$
While the fine details of the comparison will change for different backgrounds, the main features will go through.
For example, we expect our method to work for the BTZ black hole \cite{Banados:1992wn}, for which the retarded Green's function 
was computed in \cite{Son:2002sd} and \cite{Iqbal:2009fd}, as well as for the warped
$AdS_3$ black hole solutions constructed in \cite{Anninos:2008fx} for which the real time 
two-point function was computed in \cite{Chen:2009cg}.
It should also work for generic extremal black holes, for which the near horizon 
geometry was shown to include an $AdS_2$ component in \cite{Kunduri:2007vf}, as well as for higher-dimensional 
generalizations \cite{Lu:2008jk}.
In all of these cases,  three-point functions should be easily computable
applying the prescription presented herein.
While our calculation of extremal correlators is exact, our method can also be used to obtain an approximate expression
for the three-point function for generic conformal weights.
We anticipate that such an approximation should be able to capture the essential features of the 
corresponding dual conformal field theory correlators.

A calculation of the renormalized three-point correlation function using holographic renormalization 
is an interesting question to which we hope to return to in the near future
\cite{lastcit}.
Finally, we should emphasize that very little is known about specific properties of the CFT dual to Kerr.
As we have seen, there is a remarkable similarity between our results and those found  
in standard AdS/CFT studies of extremal correlators. It would be interesting to ask whether one can extract any new 
information on the structure of the CFT dual to Kerr, from the lessons learned from the string-theoretic AdS/CFT results.
For the case of Kerr black holes, and in particular the setup studied in this paper, any potential overlap would clearly be
with the work on $AdS_3$ string theory compactifications.

\acknowledgments
We would like to thank K. Becker, A. Castro, G. Compere, D. Freedman, K. Skenderis, M. Taylor, D.T. Son, A. Strominger
and C. Warnick for useful discussions. MB and WS would like to thank DAMPT (Cambridge) for the kind hospitality
during the final stages of this project.
Long walks through the beautiful Grantchester meadows provided additional inspiration.
The work of MB and WS was supported by NSF under PHY-0505757 and the University of Texas A\&M.
SC has been supported by the Cambridge-Mitchell Collaboration in Theoretical Cosmology and the Mitchell Family Foundation.

\end{document}